%Paper: hep-th/9503028
%From: Hidetoshi Awata <awata@kurims.kyoto-u.ac.jp>
%Date: Mon, 06 Mar 1995 12:04:48 JST
%Date (revised): Mon, 06 Mar 1995 14:42:57 JST
%Date (revised): Wed, 08 Mar 1995 02:33:39 JST

%%%%%%%%%%%%%%%%%%%%%%%%%%%%%%%%%%%%%%%%%%%%%%%%%%%%%%%%%%%%%%%%%%%%%%%
%
%	A Note on Calogero-Sutherland Model, $W_n$ Singular Vectors
%
%		and Generalized Matrix Models
%
%	(Based on the talk in the work shop at YITP on Dec. 6-9, 1994.)
%
%%%%%%%%%%%%%%%%%% Style %%%%%%%%%%%%%%%%%%%%%%%%%%%%%%%%%%%%%%%%%%

\magnification=\magstep1
\voffset .6in	\hoffset .5in
\vsize 6.25in	\hsize 4.6in
\baselineskip=12pt
\parskip 0pt plus 1pt	\parindent 20pt

%%%%%%%%%%%%%%%%%% foot note %%%%%%%%%%%%%%%%%%%%%%%%%%%%%%%%%%%%

%\font\srm=cmr9		\font\sit=cmti9
\font\srm=cmr10		\font\sit=cmti10
\def\foot{\baselineskip=7pt\srm\footnote}

%%%%%%%%%%%%%%%%% Definitions %%%%%%%%%%%%%%%%%%%%%%%%%%%%%%%%%%%%

\def\la{\lambda}
\def\cN{{\cal N}}	\def\cG{{\cal G}}	\def\cO{{\cal O}}

%%%%%%%%%%%%%%%%%%%%%%%%%%%%%%%%%%%%%%%%%%%%%%%%%

\def\no{\noindent}	
\def\cl{\centerline}

\def\<{\langle}	\def\>{\rangle}
\def\({\left(}	\def\){\right)}
\def\der#1{{\partial\over\partial #1}}

\def\Sect#1{\vskip4mm\no{\bf #1}\vskip2mm}
\def\Sub#1{\vskip4pt\no{\bf #1.}~}

\def\QED{\ \ \ \ \hbox{\hfill\vbox{\hrule width8pt
\noindent\vrule height8pt\hskip 7.5pt\vrule height8pt\hrule width8pt}}}

%%%%%%%% definitions only for this article %%%%%%%%%%%

\def\Inner#1#2{#1\cdot#2}
\def\va{\vec a}		\def\vQ{\vec Q}		\def\vh{\vec h}
\def\val{\vec\alpha}	\def\vLa{\vec\Lambda}	\def\vp{\vec\phi}

\def\x#1#2{t^{(#1)}_{#2}}	\def\xx#1{t^{(#1)}}
\def\t#1#2{t^{(#1)}_{#2}}
\def\gg#1#2{g^{(#1)}_{#2}}
\def\rr#1{r_{#1}}		\def\ss#1{s_{#1}}
\def\nn{N}			\def\wn{n}

\def\Pder#1{\partial_{#1}} %{{#1\over\beta}{\partial\over\partial P_#1}}
\def\Exp#1{\exp\left\{#1\right\} } %\def\Exp#1{e^{#1}} %

%%%%%%%%%%%%%%%%%%%%%%%%%%%%%%%%%%%%%%%%%%%%%%%%%%%%%%%%%%%%%%%%%%%%%%%

%%%%%%%%%%%%%%%%%%%%%%%%%%%%%%%%%%%%%%%%%%%%%%%%%%%%%%%%%%%%%%%%%%%%%%%%%
%
%                        Young diagrams
%
%%%%%%%%%%%%%%%%%%%%%%%%%%%%%%%%%%%%%%%%%%%%%%%%%%%%%%%%%%%%%%%%%%%%%%%%%
%
%%%%%%%% BoxLine %%%%%%%%%
\def\boxline#1{\vbox{\hrule\hbox{\vrule\vbox{#1}\vrule}\hrule}}
\def\boxNW#1{\vbox{\hrule\hbox{\vrule\vbox{#1}}}}
\def\boxES#1{\vbox{\hbox{\vbox{#1}\vrule}\hrule}}
%
%%%%%%% Beteen %%%%%%%%%%
\def\Between#1#2#3#4{ %% #1#2 : size, #3 : place, #4 : letter
\raise-#1mm\vbox to#1mm{\hsize #2mm \vbox{\vskip #3mm\cl{#4} } }
}
%%%%%%% small Young %%%%
\def\Young#1#2#3#4#5#6#7#8#9{
%% #1#6 : size, #5 = #1/5, #7 = #6/5, #2 = 4*#5, #3 = 3*#5, #4 = 2*#5,
%% #8 : place of \lambda, #9 = #1 - 1mm,
\raise-#9mm\boxNW{\vbox to#1mm{\hsize#6mm
	\vbox{\vskip#8mm\no\hskip#8mm$\;\lambda$} }}
\kern-#6mm
\raise-#1mm\boxES{\vbox to #5mm{\hsize #7mm $ $}}\kern-.4pt
\raise-#2mm\boxES{\vbox to #5mm{\hsize #7mm $ $}}\kern-.4pt
\raise-#3mm\boxES{\vbox to #5mm{\hsize #7mm $ $}}\kern-.4pt
\raise-#4mm\boxES{\vbox to #5mm{\hsize #7mm $ $}}\kern-.4pt
\raise-#5mm\boxES{\vbox to #5mm{\hsize #7mm $ $}}
}
%%%%%% Square Young %%%%%%%%
\def\Square#1#2#3#4{ %% #1 : size, #2 = #1/2 - 6pt, #3 = r, #4 = s,
\raise-#1mm\boxline{
\vbox to#1mm{\hsize #1mm \vbox{\vskip #2mm\no\hskip4pt {#3}
			       \vskip-#2mm\vskip-8pt\cl{#4} }} }
}
%%%%%%%%%%% Galilei %%%%%%%%%%%%
\def\Galilei{
\Between{10}{15}3{${\cal G}_s\; :\;$}
  \Young{10}8642{15}32{9.9}
\Between{10}{20}3{$\longmapsto$}
 \Square{14}7{$r$}{$s$} \kern-.4pt  % 12 & 6,  15 & 7
  \Young{10}8642{15}32{9.9}
\Between{10}{10}{10}{\quad .}
}
%%%%%%%%%% general Young %%%%%%%%%%%%%%%%
\def\Dummy{{{ }\over{ }}}
\def\generalYoung{
\Between{10}{15}5{$\lambda=$}
 \Square{15}8{$r_{n-1}$}{$\;s_{n-1}\Dummy$} \kern-.4pt
 \Square{13}7{$r_{n-2}$}{$\;\;s_{n-2}\Dummy$} \kern-.4pt
\Between{10}{13}3{$\cdots$}
 \Square{11}6{$r_2    $}{$\;\;s_2  \Dummy$} \kern-.4pt
 \Square{ 9}5{$r_1    $}{$\;\;s_1  \Dummy$} \kern-.4pt
\Between{15}{10}{10}{\hfill .}
}
%%%%%%% Big general Young %%%%%%%%%%
\def\generalYoungB{
\Between{10}{15}7{$\lambda=$}
 \Square{20}9{$r_{n-1}$}{$s_{n-1}\Dummy$} \kern-.4pt
 \Square{17}8{$r_{n-2}$}{$s_{n-2}\Dummy$} \kern-.4pt
\Between{10}{15}7{$\cdots\cdots$}
 \Square{13}6{$r_2    $}{$\;s_2  \Dummy$} \kern-.4pt
 \Square{10}5{$r_1    $}{$\;s_1  \Dummy$} \kern-.4pt
\Between{15}{10}{10}{\hfill .}
}
%%%%%%%%%%%%%%%%%%%%%% end of figures %%%%%%%%%%%%%%%%%%%%%%%%%%%%%%%%%

%%%%%%%%%%%%%%%%%%%%%%%%%%%%%%%%%%%%%%%%%%%%%%%%%%%%%%%%%%%%%%%%%%%%%%%%%
%
%                        References
%
%%%%%%%%%%%%%%%%%%%%%%%%%%%%%%%%%%%%%%%%%%%%%%%%%%%%%%%%%%%%%%%%%%%%%%%%%

%%%%%%%%%%%%%%%%%% reference list %%%%%%%%%%%%%%%%%%%%%%%%%%%%%

\def\RfAMOS{H.~Awata, Y.~Matsuo, S.~Odake and J.~Shiraishi,
	{\sit Collective Field Theory, Calogero-Sutherland Model
	and Generalized Matrix Models},
	preprint, hep-th/9411053, to appear in Phys. Lett. {\bf B}; %\hb
	{\sit Excited states of Calogero-Sutherland model and
	singular vectors of the $W_N$ algebra},
	to appear}

\def\RfFL{V. Fateev and S. Lykyanov,
	Int. J. Mod. Phys. {\bf A3} (1988) 507--520}

\def\RfH{Z.N.C.~Ha, Phys. Rev. Lett. {\bf 60} (1994) 1574--1577 ;
	Nucl. Phys. {\bf B435} (1995) 604--636} %cond-mat/9405063

\def\RfHar{G.R. Harris,	Nucl. Phys. {\bf B} (1991) 685--702}

\def\RfIK{N.~Ishibashi and H.~Kawai, Phys. Lett. {\bf B314} (1993) 190--196}

\def\RfJS{A.~Jevicki and B.~Sakita, Nucl. Phys. {\bf B165} (1980) 511-527}

\def\RfKMMM{S. Kharchev, A. Marshakov, A. Mironov and A. Morozov,
	Nucl. Phys. {\bf B404} (1993) 717--750}

\def\RfLPS{F.~Lesage, V.~Pasquier and D.~Serban,
	Nucl. Phys. {\bf B435} (1995) 585--603} %hepth/9405008

\def\RfM{I. Macdonald, Publ. I.R.M.A. (1988) 131--171}

\def\RfMP{J.A.~Minahan and A.P.~Polychronakos,
  	{\sit Density Correlation Functions in Calogero--Sutherland Models},
  	preprint, hep-th/9404192} %hepth/9404192

\def\RfMYi{K.~Mimachi and Y.~Yamada,
  	{\sit Singular vectors of the Virasoro algebra in terms of
  	Jack symmetric polynomials},
  	preprint (November 1994)}

\def\RfMYii{Similar result was claimed by K. Mimachi and Y. Yamada}

\def\RfSt{R. Stanley, Adv. Math. {\bf 77} (1989) 76--115}

\def\RfSu{B.~Sutherland,
	Phys. Rev. {\bf A4} (1971) 2019--2021; {\bf A5} (1992) 1372--1376}
	%Jour. Math. Phys. {\bf 12} (1970) 246-250, 251-256}

%%%%%%% reference number %%%%%%%

\def\ref#1{$^{#1}$}

\def\rSu	{1}
\def\rH		{2}
\def\rLPS	{3}
\def\rMP	{4}
\def\rAMOS	{5}
\def\rMYii	{6}
\def\rMYi	{7}
\def\rM		{8}
\def\rSt	{9}
\def\rFL	{10}
\def\rJS	{11}
\def\rIK	{12}
\def\rHar	{13}
\def\rKMMM	{14}

%%%%%%%% reference out %%%%%%%%%

\def\RefOut{
{\baselineskip9pt
\srm
\item{[\rSu	]}{\RfSu	}
\item{[\rH	]}{\RfH		}
\item{[\rLPS	]}{\RfLPS	}
\item{[\rMP	]}{\RfMP	}
\item{[\rAMOS	]}{\RfAMOS	}
\item{[\rMYii	]}{\RfMYii	}
\item{[\rMYi	]}{\RfMYi	}
\item{[\rM	]}{\RfM		}
\item{[\rSt	]}{\RfSt	}
\item{[\rFL	]}{\RfFL	}
\item{[\rJS	]}{\RfJS	}
\item{[\rIK	]}{\RfIK	}
\item{[\rHar	]}{\RfHar	}
\item{[\rKMMM	]}{\RfKMMM	}
\item{ }{ }
}}

%%%%%%%%%%%%%%%%%% end of references %%%%%%%%%%%%%%%%%%%%%%%%%%%%%%%%%

%%%%%%%%%%%%%%%%%%%%%%%%%%%%%%%%%%%%%%%%%%%%%%%%%%%%%%%%%%%%%%%%%%%%%%%%
%
%		Tytle
%
%%%%%%%%%%%%%%%%%%%%%%%%%%%%%%%%%%%%%%%%%%%%%%%%%%%%%%%%%%%%%%%%%%%%%%%%

\def\talk{Based on the talk in the work shop at YITP on Dec. 6--9, 1994.}

\rightline{hep-th/9503028}\vskip5pt
\cl{\bf A Note on Calogero-Sutherland Model, $W_\wn$ Singular Vectors}
\cl{\bf and Generalized Matrix Models{\foot{$^\dagger$}\talk} }

%%%%%%%%%%%%%%%%%%%%% name %%%%%%%%%%%%%%%%%%%%%%%%%%%%%%%%%%%%%%%%%%

\def\JSPS{JSPS fellow}
\vskip9pt
\cl{
H. Awata$^1${\foot{$^\ddagger$}\JSPS},
Y. Matsuo$^2$,
S. Odake$^3$ and
J. Shiraishi$^{4\ddagger}$
}

\vskip6pt
{\baselineskip9pt
\srm
\cl{$^1$ Research Institute for Mathematical Sciences}
\cl{Kyoto University, Kyoto 606, Japan}
\vskip1.5pt
\cl{$^2$ Uji Research Center, Yukawa Institute for Theoretical Physics}
\cl{Kyoto University, Uji 611, Japan}
\vskip1.5pt
\cl{$^3$ Department of Physics, Faculty of Liberal Arts}
\cl{Shinshu University, Matsumoto 390, Japan}
\vskip1.5pt
\cl{$^4$ Department of Physics, Faculty of Science}
\cl{University of Tokyo, Tokyo 113, Japan}
}

%%%%%%%%%%%%%%%%%%%%%%%%%%%%%%%%%%%%%%%%%%%%%%%%%%%%%%%%%%%%%%%%%
\vskip9pt
\cl{\bf Abstract}
\vskip2pt
%%%%%%%%%%%%%%%%%%%%%%%%%%%%%%%%%%%%%%%%%%%%%%%%%%%%%%%%%%%%%%%%%

We review some recent results on the Calogero-Sutherland model
with emphasis upon its algebraic aspects.
We give integral formulae for excited states (Jack polynomials)
of this model
and their relations with $W_n$ singular vectors and generalized matrix models.

%%%%%%%%%%%%%%%%%%%%%%%%%%%%%%%%%%%%%%%%%%%%%%%%%%%%%%%%%%%%%%%%%

\Sect{ 1. Introduction}

%%%%%%%%%%%%%%%%%%%%%%%%%%%%%%%%%%%%%%%%%%%%%%%%%%%%%%%%%%%%%%%%%%
\no
The Calogero-Sutherland models\ref{\rSu} is
a quantum mechanical system
with a long-range interaction.
It has been actively studied as a
 solvable system with anyonic statistics in $1+1$ dimensions.
%
%%%%%%%%%%%% dinamical CF %%%%%%%%%%
%
Recently, it is greatly developed with the calculations
of dynamical correlation functions\ref{\rH, \rLPS, \rMP}.
To evaluate more general ones,
we may need to express
the wave-functions of the excited states explicitly.
In this note,
we will present their algebraic construction
by integral transformations\ref{\rAMOS, \rMYii}.

%%%%%%%%% relation with other theories %%%%%%%%%%%%%%

For a special value of the coupling constant,
the system reduces to that of free fermion.
In this case, it has deep relations with
the $W$ algebras, matrix models,
2D quantum gravities and also 2D QCD.
Such connections should remain even for the general cases.
Indeed, we will demonstrate that
the wave-functions of excited states are
identified with the $W_\wn$ singular vectors\ref{\rAMOS, \rMYi}.
Furthermore, the $W_\wn$ structure thus obtained
causes the $W$ constraints in the generalized matrix models.

%\vfill\eject
%%%%%%%%%%%%%%%%%%%%%%%%%%%%%%%%%%%%%%%%%%%%%%%%%%%%%%%%%%%%%%%%%

\Sect{ 2. Calogero-Sutherland model and Jack polynomial.}

%%%%%%%%%%%%%%%%%%%%%%%%%%%%%%%%%%%%%%%%%%%%%%%%%%%%%%%%%%%%%%%%%%
\no
We start with recapitulating some important properties of
the Calogero-Sutherland Hamiltonian and momentum.

\Sub{2.1}
We consider a $\nn$--body problem on a unit circle.
Denote their coordinates $q_1,q_2,\cdots, q_\nn$.
Hamiltonian and momentum are given by
$$
H_{CS}=\sum_{j=1}^\nn \({1\over i}\der{q_j}\)^2
  +{1\over 2}\sum_{i<j}
{\beta (\beta -1) \over  \sin^2 (q_i-q_j)/2  },\qquad
P = \sum_{j=1}^\nn {1\over i}\der{q_j}.
\eqno{(2.1)}$$
Here $\beta$ is a coupling constant
with a reflection symmetry $\beta \leftrightarrow 1-\beta$.
It is known that  when $\beta$ is a real number,
i.e., $\beta (\beta -1)\geq -1/4$,
the system becomes stable and has no bound states.

%%%%%%% inner product %%%%%%

Define an inner-product
$
\(f,g\)\equiv\int_0^{2\pi}\prod_j dq_j\, f(q)^*\, g(q)
$.
Notice that $H_{CS}$ and $P$ are self-adjoint
$H_{CS}^\dagger = H_{CS}$ and $P^\dagger = P$
under $\dagger$ defined by
$\(f,\cO\; g\) \equiv \(\cO^\dagger f,g\)$.

%%%%%%%%%%%%% vacuum %%%%%%%%%%%%%

Since the Hamiltonian is rewritten as
$$\eqalign{
H_{CS} &= \sum_j h_j(\beta )^\dagger\, h_j(\beta )+\varepsilon_0,\cr
h_j(\beta )&={1\over i}{\der{q_j}}
+\beta\sum_{i(\neq j)}\cot\({q_j-q_i\over 2}\),
}$$
with a vacuum energy $\varepsilon_0$,
the energy is bounded from below and
there are two minimal-energy states characterized as
the states annihilated by $h_j(\beta )$ or $h_j(1-\beta )$.
We restrict ourselves to the former vacuum, which is
$$
\Delta^\beta \equiv \prod_{i<j} \sin^\beta \({q_i-q_j\over 2}\).
$$
The statistic of the particle is governed by the coupling $\beta $:
if $\beta$ is even (odd) then the particles become bosonic (fermionic).

%%%%%%%%%%%%%%%%%%%%%%%% excitation %%%%%%%%%%%%%%%%%%%%%%%%%%%%%%%%%%%%%%

\Sub{2.2}
We will write the excited states
in the factorized form $J(q)\Delta (q)^\beta $ and
change the variables to $x_j\equiv\exp(iq_j)$ on a complex plane.
In these new variables, $J(x)$ has to be a symmetric function,
to possess the same statistic as the vacuum.
Define new Hamiltonian and momentum
acted directly on $J(x)$ as follows:
$H \equiv\Delta^{-\beta }H_{CS}\Delta^\beta -\varepsilon_0$ and
$P\equiv\Delta^{-\beta }P\Delta^\beta$, then
$$
H =
\sum_{i=1}^\nn D_i^2 +
\beta \sum_{i<j} { x_i+x_j \over  x_i-x_j }(D_i-D_j),\qquad
P= \sum_{i=1}^\nn D_i,
\eqno{(2.2)}$$
with $D_i = x_i \der{x_i}$.
The vacuum wave-function becomes
$\Delta^\beta = \prod_{i\neq j}(1-x_i/x_j)^{\beta /2}$.

%%%%%%%%%%% Jack %%%%%%%%%%%%%

The eigenstate is labeled by a decreasing set of non-negative integers,
$\la=(\la_1\geq\la_2\geq\cdots\geq\la_\nn\geq 0)$,
which is identified with a Young diagram
with $\la_i(\geq 1)$ squares in $i$th row.
The corresponding eigenvalue of $H_{CS}$ and $P$ can be written
in terms of the momenta $k_j$ of pseudo-particles as
$$
\varepsilon_\la= \sum_{i=1}^N k_i^2,\qquad
p_\la= \sum_{i=1}^N k_i,\qquad
k_j\equiv\la_j + {\beta\over 2} (\nn+1-2j),
$$
respectively.
The wave-function $J_\la(x)$ is called the Jack symmetric polynomial
in mathematical literatures\ref{\rSu, \rM, \rSt}.
The neighboring pseudo-momenta should satisfy,
$k_i-k_{i+1}\geq \beta$, which exhibits
the nature of fractional statistics of this system.
When $\beta =1$,
the Jack polynomial reduces to the Schur polynomial.

%%%%%%%%%%% pseudo-momentum %%%%%%%%%%%%%%%%

\def\Pseudo{
It may be instructive to illustrate the spectrum of the system for
a positive integer $\beta $.
We fill the momentum occupied by
the pseudo-particle with ``$1$'' and the vacant state with ``$0$'' in
the integer-valued momentum space.
We denote the origin ($p=0$) by ``$\, : \,$''.
For example, the $4$--particles and $\beta =3$ case,
$(1)$ the vacuum; and
$(2)$ the excited state with $\la=(3,1,1,0)$ are respectively
%$$\eqalign{&(1),\qquad\cdots\;\;0\;\;1\;\;0\;\;0\;\;1\;\;0 :
%0\;\;1\;\;0\;\;0\;\;1\;\;0\;\;0\;\;0\;\;0\;\;0\;\;0\;\;0\;\;\cdots,
%\qquad\qquad \cr
%&(2),\qquad\cdots\;\;0\;\;1\;\;0\;\;0\;\;0\;\;0 :
%1\;\;0\;\;0\;\;1\;\;0\;\;0\;\;0\;\;0\;\;0\;\;0\;\;1\;\;0\;\;\cdots.
%\qquad\qquad}$$
$$\eqalign{
&(1),\qquad\cdots
\;\;0\;\;1\;\;0\;\;0\;\;1\;\;0 :
0\;\;1\;\;0\;\;0\;\;1\;\;0\;\;0\;\;0\;\;0\;\;\cdots,\qquad\qquad \cr
&(2),\qquad\cdots
\;\;0\;\;1\;\;0\;\;0\;\;0\;\;1 :
0\;\;0\;\;1\;\;0\;\;0\;\;0\;\;0\;\;1\;\;0\;\;\cdots.\qquad\qquad
}$$
}
\Pseudo

%%%%%%%%%%%%%%%%%%%% power-sum Hamiltonian %%%%%%%%%%%%%%%%%%%%%%%%%%%%%%

\Sub{2.3}
Since $H$ and $P$ are symmetric in $x_i$'s,
they can be expressed by the power-sums $p_n\equiv\sum_{i=1}^\nn x_i^n$
and their derivatives ${\Pder n} \equiv {n\over\beta }\der{p_n}$
as follows\ref{\rSt}:
$$
H = \beta^2 \sum_{n,m>0} %{n,m=1}^\nn
\(p_{n+m} {\Pder n} {\Pder m} + p_n \,p_m {\Pder {n+m}} \)
+ \beta \sum_{n>0} %{n=1}^\nn
(n-n\beta +N\beta ) \,p_n {\Pder n},
$$
$$
P = \beta \sum_{n>0} %{n=1}^\nn
p_n {\Pder n}.
\eqno{(2.3)}$$\def\ePH{(2.3)}\no
Here we must treat $p_n$'s as formally independent variables, i.e.,
${\Pder n}\, p_m = {n\over\beta}\delta_{n,m}$ for all $n,m>0$.
Since this $H$ and $P$ do not depend on %include
the number $N$ of particles up to the last term of $H$,
they are useful in analizing $N$--independent properties.
%to the analysis.

%%%%%%% inner-product %%%%%%%%%%

Define another inner-product
$
\<f,g\>\equiv\oint\prod_j {dp_j\over p_j}\, \overline{f(p)}\,g(p)
$,
with $\overline{p_n}\equiv{\Pder n}$.
This is nothing but that of free bosons.
Notice that $H$ and $P$ have a duality
$p_n \leftrightarrow {\Pder n}$, i.e.,
they are self-adjoint $H^\ddagger = H$ and $P^\ddagger = P$
under $\ddagger$ defined by
$\<f,\cO\; g\> \equiv \<\cO^\ddagger f,g\>$.

%%%%% Virasoro in Hamiltonian  %%%%%%%

There exists a following relation between $H$ and
the non-relativistic Virasoro generators ${\cal L}_n$:
$$\eqalign{
&H = \beta \sum_{n>0} %{n=1}^{2\nn}
p_n \,{\cal L}_n + (\beta-1+\beta N)\,P,\cr
{\cal L}_n =
\beta \sum_{m=1}^{n-1} &{\Pder m} {\Pder{n-m}} +
\beta \sum_{m>0} %{m=1}^{N-n}
p_m {\Pder{n+m}}
 -(n+1)\(\beta -1\) {\Pder n}.
}\eqno{(2.4)}$$\def\eVH{(2.4)}\no
These relations with free bosons or Virasoro generators
suggest algebraic aspects of the model.

%%%%%%%%%%%%%%%%%%%%%%%%%%%%%%%%%%%%%%%%%%%%%%%%%%%%%%%%%%%%%%%%%

\Sect{ 3. Integral formula for the wave-functions of excited states}

%%%%%%%%%%%%%%%%%%%%%%%%%%%%%%%%%%%%%%%%%%%%%%%%%%%%%%%%%%%%%%%%%%
\no
We next try to derive the explicit expression of all excited states.
Our strategy is as follows:
we introduce two types of (integral) transformations
which maps the eigenstate into another
while changing its energy and the number of particles.
We can construct arbitrary state by applying
them successively to the vacuum.

%%%%%%%%%%%%%% Galilean boost %%%%%%%%%%%%%%%%%%%%%

First, we introduce
the Galilean boost $\cG_s$, which
uniformly shifts the pseudo-momentum of the pseudo-particles
from $\la=(\la_1,\cdots,\la_r)$ to
$\la+ s^r =(\la_1 + s,\cdots,\la_r + s)$.
It can be realized by  multiplying
the wave-function by
$\prod_j e^{i q_j s} = \prod_j x_j^s$.
When it is operated to the eigenstate,
the Young diagram is changed by adding a rectangle
$s^r$ from the left: 		%as follows
%$$\cG_s : J_\la(x)\mapsto J_{\la+s^r}(x),$$
$$
\cG_s\cdot J_\la(x_1,\cdots,x_r)=
J_{\la+s^r}(x_1,\cdots,x_r)=
\prod_{i=1}^r x_i^s\cdot J_\la(x_1,\cdots,x_r).
\eqno{(3.1)}$$
$$
\Galilei
$$
%Note that $J_{s^r}(x_1,\cdots,x_r)=\prod_{i=1}^r x_i^s$.

%%%%%%%%%%% particle number changing %%%%%%%%%%%%

The second integral transformation $\cN_{NM}$
changes the number of particles from $M$ to $N$: 	%as follows:
%$$\cN_{NM} : J_\la(t_1,\cdots,t_M)\mapsto J_\la(x_1,\cdots,x_N),$$
$$\eqalign{
\cN_{NM}&\cdot J_\la(t_1,\cdots,t_M) =J_\la(x_1,\cdots,x_N)\cr
&=
\oint\prod_{j=1}^M {dt_j\over t_j}
\prod_{i,j}\(1-{x_i/t_j}\)^{-\beta}
\prod_{i\neq j}^M \(1-{t_i/t_j}\)^\beta
J_\la(t_1,\cdots,t_M),
}\eqno{(3.2)}$$\def\eSecondTrans{(3.2)}
where the integration path is along
the unit circle in the complex plane.
%where the integral $\oint{dt/ t}$ stands for
%$\int_0^{2\pi}dq$ with $t = e^{iq}$.

%%%%%%%% self-dualities %%%%%%%%%%
%
\def\degeneracy{
Since the energy degenerates,
we need one more condition to define the Jack polynomial uniquely.
However, $\eSecondTrans$ is also compatible with it{\ref\rAMOS}.
}
\no{\it Proof.}
This is proved by using two self-dualities 		%adjointnesses
of Hamiltonians and momentum. 			%as follows.
%
%%%%%%%%%% commutativity %%%%%%%%%%%%
%
Replace the variables $t_j$ of integration with $t_j^{-1}$. 	%$1/t_j$.
Let
$$\eqalign{
&\widetilde H(x)\equiv\( H(x) - \beta NP(x)\),\cr
V\equiv\prod_{i,j}&\(1-{x_i t_j}\)^{-\beta }
 = e^{\beta\sum_{n>0}{1\over n} \sum_{i,j} x_i^n t_j^n}.
% = {\Exp{\sum_{n>0}\sum_{i,j}{\beta \over n} x_i^n t_j^n}}.
}$$
Then $\widetilde H$ commutes with $\cN_{NM}$ as
$$
\widetilde H(x_1,\cdots,x_N) \, \cN_{NM}
= \cN_{NM} \,\widetilde H(t_1,\cdots,t_M),
$$
which is deduced as follows:
first we change the action with
$\widetilde H(x)$ on $V$
to that with $\widetilde H(t)$ as
$\widetilde H(x) V = \widetilde H(t)^\ddagger V = \widetilde H(t) V$;
next we perform the integration by parts and
pass the Hamiltonian through $\Delta^{2\beta}$ as
$\Delta^{-2\beta }H^\dagger\Delta^{2\beta}
=\Delta^{-\beta }H_{CS}^\dagger\Delta^\beta -\varepsilon_0 = H$.
Momentum $P$ also commutes with it{\foot{$^\star$}\degeneracy}.
\QED

%%%%%%%%%%%%% integral formula for Jack %%%%%%%%%%%%%

Therefore, starting from the vacuum with ${\ss 1}$ particles
and combining these two transformations $\cG_s$ and $\cN_{nm}$,
we obtain all excited states of $\nn$ particles\ref{\rAMOS,\rMYii}
$$\eqalign{
&J_\la(x)
= \cN_{{\rr\wn},{\rr{\wn-1}} } \cG_{\ss{\wn-1}}
  \cN_{{\rr{\wn-1}},{\rr{\wn-2}} }
  \cdots\cdots
  \cG_{\ss 2} \cN_{{\rr 2},{\rr 1} } \cG_{\ss 1}\cdot 1 \cr
&
= \oint\prod_{a=1}^{\wn-1}\prod_j^{\rr a} {{d\x aj}\over{\x aj}}
\prod_{i,j} \(1- {\x {a+1}i}/{\x aj}\)^{-\beta}
\prod_{i\neq j}^{\rr a} \(1-{\x ai}/{\x aj}\)^\beta
\prod_{j=1}^{\rr a}\({\x aj}\)^{\ss a},
}\eqno{(3.3)}$$\def\eIJ{(3.3)}\no
with $x_j\equiv{\x \wn j}$, $\nn=\rr\wn$ and
$\la=\sum_{a=1}^{\wn-1}\({\ss a}^{\rr a}\)$ such that
$$
\generalYoung
$$
%
%The normalization of the eigenfunction is derived in\ref{\rAMOS}.
This formula reveals new algebraic aspects of the model
as the following two sections.

%%%%%%%%%%%%%%%%%%%%%%%%%%%%%%%%%%%%%%%%%%%%%%%%%%%%%%%%%%%%%%%%%

\Sect{ 4. Relation with $W_\wn$ singular vectors
and WZNW correlation functions}

%%%%%%%%%%%%%%%%%%%%%%%%%%%%%%%%%%%%%%%%%%%%%%%%%%%%%%%%%%%%%%%%%%
\no
The integrand of eq.\ $\eIJ$ reminds us of a $sl(\wn)$ type chain
from ${\xx 1}$ to ${\xx {\wn-1}}$.
Indeed, they are realized by the $sl(\wn)$ type boson $\vp(z)$ such that
$$\eqalign{
&{\Inner{\val^a}{\vp}}(z)\; {\Inner{\val^b}{\vp}}(w)
\sim A^{ab}\log(z-w),\cr
\vp(&z)\equiv -\sum_{m\neq 0}{1\over m}\va_m z^{-m} + \va_0 \log z + \vQ,
}$$
with simple roots $\val^1,\cdots,\val^{\wn-1}$
and the $sl(\wn)$ type Cartan matrix $A^{ab}$. \break %\hb
Through this correspondence,
we show that the Jack polynomials are
identified with $W_\wn$ singular vectors
after a projection defined below.

%%%%%%%%%% singular vectors %%%%%%%%%%%%%%%%%%%

Let us consider the bosonic Fock space
generated by the highest weight state $|\vh\>$ such that
${\Inner{\val^a}{\va_0}}|\vh\> = h^a |\vh\>$ and
${\va_m}|\vh\> = 0$ $(m>0)$.
For the $W_\wn$ algebra
with a Virasoro central charge
$$
c=(\wn-1)\left\{1-\wn(\wn+1)\(\sqrt\beta -{1\over\sqrt\beta}\)^2\right\},
$$
there exists a singular vector on the Fock space of the highest weight
$$
h_{r,s}^{\wn-a} =
(1+{\rr a}-{\rr{a+1}})\sqrt\beta -(1+{\ss a}){1\over\sqrt\beta},
$$
with positive-integers ${\rr a}<{\rr{a+1}}$ and ${\ss a}$.
Its Virasoro grade is $\sum_{a=1}^{\wn-1} {\rr a}{\ss a}$.
It is constructed from the screening currents
$: {\Exp{\sqrt\beta {\Inner{\val^a}{\vp}}(t)}}:$ as follows\ref{\rFL}
$$
|\chi_{r,s}\> = \oint\prod_{a,j}d{\t aj}
: e^{\sqrt\beta {\Inner{\val^a}{\vp}}({\t aj})}: |\vh_{r,s}^\prime\>.
\eqno{(4.1)}$$
If we perform a OPE of this singular vector,
then the OPE factor is almost the same as
the integrand of eq.\ $\eIJ$.
In fact, the excited state and the singular vector relate
as follows\ref{\rAMOS} (ref.\ [\rMYi] for the Virasoro case):
$$
J_\la(x) = \<\vh_{r,s}|\,V_1\,|\chi_{r,s}\>.
\eqno{(4.2)}$$
Here
$$
\<\vh |V_1\equiv
\<\vh |e^{\sqrt\beta \sum_{m>0}{1\over m}{\Inner{\vLa_1}{\va_m}}\,p_m} ,
$$
with fundamental weights $\vLa_a$ such that
${\Inner{\val^a}{\vLa_b}}=\delta^a_b$.
It gives a projection from $\wn-1$ bosons to power-sums as
$$
\<\vh |V_1\,{\Inner{\val^a}{\va_{-m}}}
=\delta_1^a\sqrt\beta p_m\<\vh |V_1,\qquad
\<\vh |V_1\,{\Inner{\vLa_1}{\va_m}}
={m\over\sqrt\beta }\der{p_m}\<\vh |V_1,
$$
for a positive integer $m$.
%and $|\alpha\> \rightarrow 1$, after performing a OPE.

%%%%%%%% Cubic-Hamiltonian with bosons %%%%%%%%%%%%%%%%

By using eq.\ $\ePH$ and the above projection,
one can consider a bosonic Hamiltonian $\widehat H$
acted directly on the bosonic Fock space.
Although it is not uniquely detemined,
the nontrivial part of $\widehat H$ is a cubic form\ref{\rJS}
similar to Ishibashi-Kawai Hamiltonian of string fields\ref{\rIK}.
Furthermore,
$\widehat H$ is expressed by using Virasoro generators $L_m$ as $\eVH$:
$$
\widehat H \sim \sum_{m>0} {\Inner{\val^1}{\va_{-m}}}\,L_m + \cdots,
$$
which has a similaritiy with the BRST-operator
of the two dimensional quantum gravity.
Here $\cdots$ are the Cartan parts
and the trivial ones that are annihilated by the projection.
Therefore, we obtain another view point for the integral formula $\eIJ$:
the $W_\wn$ singular vectors in terms of bosons always
become the eigenstates of the Calogero-Sutherland model
because they are annihilated by the cubic part of the Hamiltonian
$\widehat H$.

%%%%%%%%%%%%%% KZ %%%%%%%%%%%%%%%%%%%%%%%%%

There is also a relation with the correlation function of the WZNW model.
Let us consider the Kac-Moody algebra $\widehat{sl(\wn)}$
with a level $\kappa\equiv k+\wn$.
The vertex operator $V_1$ in the projection operator
corresponds to the product
$\prod_{i=1}^\nn :{\Exp{\sqrt{1/\kappa}{\Inner{\vLa_1}{\vp}}(x_i)} }:$
of $\nn$--vertex operators of $\widehat{sl(\wn)}$
with fundamental representations.
Furthermore, the screening current of the $W_\wn$ algebra
is nothing but the $\phi$--part (without $\beta \gamma $--part)
$:{\Exp{-\sqrt{1/\kappa}{\Inner{\val^a}{\vp}}(t)} }:$
of that of $\widehat{sl(\wn)}$.
If we decompose the Young diagram as ${\rr a} = a$
and allow ${\ss a}$'s to vanish,
then $\nn=\wn$ and the integrand of $\eIJ$
is just the $\phi$--part of the integral formula for
a weight zero $\nn$--point function
with fundamental representations of the WZNW model
up to a non-symmetric part.

\vfill\eject
%%%%%%%%%%%%%%%%%%%%%%%%%%%%%%%%%%%%%%%%%%%%%%%%%%%%%%%%%%%%%%%%%

\Sect{ 5. Generalized matrix model and Virasoro constraint}

%%%%%%%%%%%%%%%%%%%%%%%%%%%%%%%%%%%%%%%%%%%%%%%%%%%%%%%%%%%%%%%%%%
\no
The excited state $\eIJ$ has also some similarity  to
the partition function of the matrix model.
Indeed, when $\wn=2$ and ${\ss 1}=\b(1-{\rr 1})-1$,
which is no longer a positive integers in general,
then $Z_1(g)\equiv J_\la(x)$ is
$$
Z_1(g)
= \int \prod_{i=1}^{\rr 1} dt_i
  \prod_{i<j}(t_i-t_j)^{2\beta }
  e^{\sum_{n>0} \sum_{i=1}^{\rr 1} g_n t_i^n},
\eqno{(5.1)}$$
with $g_n\equiv {\beta \over n}p_n$.
The orthogonal, hermitian and symplectic matrix models
correspond to when $\beta = 1/2$, $1$ and $2$, respectively.

%%%%%%%%%% Vir. consttaint %%%%%%%%%%%%%%%%%

Moreover,
since the Jack polynomial is constructed from the screening current,
this integral also satisfies Virasoro constraint
${\cal L}_m(g)\,Z_1(g) = 0$ for $m=-1,0,\cdots$.
This Virasoro generator is that of $\wn=2$ in the last section
with a central charge $c=1-6\(\sqrt\beta - 1/\sqrt\beta \)^2$.
Hence, this integral is a generalization of one-matrix model
for a general coupling constant $\beta $ \ref{\rHar}.

%%%%%%%%% generalized conformal matrix model %%%%%%%%%%%

Although our correspondence between bosons and power-sums is
a projection unless the Virasoro case,
one can make it an invertible map
by introducing many kinds of power-sums.
In fact, the operator
$$
\<\vh|V_{\wn-1}\equiv\<\vh|\prod_{b=1}^{\wn-1}
e^{\sqrt\beta \sum_{m>0}{1\over m}{\Inner{\vLa_b}{\va_m}}\,{\gg bm}},
$$
gives that from bosons to
$\wn-1$ kinds of power-sums ${\gg an}$ by
$$
\<\vh|V_{\wn-1}\,{\Inner{\val^a}{\va_{-m}}}
=\sqrt\beta {\gg am}\<\vh|V_{\wn-1},\quad
\<\vh|V_{\wn-1}\,{\Inner{\vLa_a}{\va_m}}
={m\over\sqrt\beta }\der{\gg am}\<\vh|V_{\wn-1},
$$
for a positive integer $m$.
When ${\ss a}=\beta (1-{\rr a}+{\rr {a+1}})-1$ with ${\rr \wn}=0$,
$$\eqalign{
& Z_{\wn-1}(g) \equiv \<\vh_{r,s}|\,V_{\wn-1}\,|\chi_{r,s}\> \cr
&= \int\prod_{a=1}^{\wn-1} \prod_{j=1}^{\rr a} d{\t aj}
e^{\sum_{n>0}{\gg an}({\t aj})^n}
                    \prod_{i<j} ({\t ai}-{\t aj})^{2\beta }
\prod_{a=1}^{\wn-2} \prod_{i,j} ({\t ai}-{\t {a+1}j})^{-\beta }.
}\eqno{(5.2)}$$
Since $Z_{\wn-1}(g)$ now saticefies a $W_\wn$ constraint,
it is regarded as a generalization of the partition function of
the conformal matrix model\ref{\rKMMM} of $\beta =1$.

%%%%%%%%%%%%%%%%%%%%%%%%%%%%%%%%%%%%%%%%%%%%%%%%%%%%%%%%%

\Sect{ References}

%%%%%%%%%%%%%%%%%%%%%%%%%%%%%%%%%%%%%%%%%%%%%%%%%%%%%%%%%%%%

\RefOut

%%%%%%%%%%%%%%%%%%%%%%%%%%%%%%%%%%%%%%%%%%%%%%%%%%%%%%%%%%%%%%%%%%

\bye

%%%%%%%%%%%%%%%%%%%%%%%%%%%%%%%%%%%%%%%%%%%%%%%%%%%%%%%%%%%%%%%%%%

%%%%%%%%%%%%%%%%%%%%%%%%%%%%%%%%%%%%%%%%%%%%%%%%%%%%%%%%%

\Sect{ References}

%%%%%%%%%%%%%%%%%%%%%%%%%%%%%%%%%%%%%%%%%%%%%%%%%%%%%%%%%%%%
{\baselineskip9pt
\srm

\item{[AMOS]}
{H.~Awata, Y.~Matsuo, S.~Odake and J.~Shiraishi,
{\sit Collective Field Theory, Calogero-Sutherland Model
   and Generalized Matrix Models},
preprint, hep-th/9411053, to appear in Phys. Lett. {\bf B}; %\hb
{\sit Excited states of Caligero-Sutherland model and
	singular vectors of the $W_N$ algebra},
to appear}

\item{[FL]}
{V. Fateev and S. Lykyanov, Int. J. Mod. Phys. {\bf A3} (1988) 507--520}

\item{[H]}{Z.N.C.~Ha,
	Phys. Rev. Lett. {\bf 60} (1994) 1574--1577 ;
	Nucl. Phys. {\bf B435} (1995) 604--636} %cond-mat/9405063

\item{[Har]}{G.R. Harris, Nucl. Phys. {\bf B} (1991) 685--702}

\item{[IK]}
{N.~Ishibashi and H.~Kawai, Phys. Lett. {\bf B314} (1993) 190--196}

\item{[JS]}
{A.~Jevicki and B.~Sakita, Nucl. Phys. {\bf B165} (1980) 511-527}

\item{[KMMM]}{S. Kharchev, A. Marshakov, A. Mironov and A. Morozov,
Nucl. Phys. {\bf B404} (1993) 717--750}

\item{[LPS]}{F.~Lesage, V.~Pasquier and D.~Serban,
Nucl. Phys. {\bf B435} (1995) 585--603} %hepth/9405008

\item{[M]}{I. Macdonald, Publ. I.R.M.A. (1988) 131--171}

\item{[MP]}{J.A.~Minahan and A.P.~Polychronakos,
  {\sit Density Correlation Functions in Calogero-Sutherland Models},
  preprint, hep-th/9404192} %hepth/9404192

\item{[MY1]}{K.~Mimachi and Y.~Yamada,
  {\sit Singlar vectors of the Virasoro algebra in terms of
  Jack symmetric polynomials},
  preprint (November 1994)}

\item{[MY2]}{Similar result was claimed by K. Mimachi and Y. Yamada}

\item{[St]}{R. Stanley, Adv. Math. {\bf 77} (1989) 76--115}

\item{[Su]}
{B.~Sutherland,
Phys. Rev. {\bf A4} (1971) 2019--2021; {\bf A5} (1992) 1372--1376}
%Jour. Math. Phys. {\bf 12} (1970) 246-250, 251-256}

}

%%%%%%%%%%%%%%%%%%%%%%%%%%%%%%%%%%%%%%%%%%%%%%%%%%%%%%%%%%%%%%%%%%

\bye